\documentclass[onecolumn]{article}
\usepackage{amssymb}
\usepackage{amsmath}

\input{tcilatex}

\begin{document}

\title{Causal transformation of G\"{o}del-type space-times in conformal
field theory}
\author{Pawel Gusin, \\
University of Silesia, Institute of Physics, ul. Uniwersytecka 4, \\
PL-40007 Katowice, Poland}
\maketitle

\begin{abstract}
The G\"{o}del-type metrices are considered as backgrounds of the
sigma-models. In the conformal field theory such backgrounds are deformed by
the exactly marginal operators. We examinate, how the closed time-like
curves (CTC's) transform under such deformations.

\begin{description}
\item[PACS] :11.25.Hf ; 04.20.Gz

\item[Keywords] : G\"{o}del-type metrices, marginal deformations.
\end{description}
\end{abstract}

\section{Introduction}

Space-times with the closed timelike curves (CTC) are not rare solutions in
general relativity. Some parts of these solutions are non-singular. However
they are not considered as physical because they violate causality principle
and lead to the well-known time paradoxes. From the other side one can
suspect that these solutions should have counterparts in quantum theory of
gravitation. This quantum theory should give a selection rule which forbids
space-times with CTC on the classical level. Hawking called this rule
chronology protection [1]. If such selection rule exists, then time
traveling in the past is impossible. In other case a time machine is
possible. Since theory of the full quantum gravity does not exist one can
only study approximately this problem. String/M-theory is such approximation
of the quantum gravity. Thus it seems interesting to examine conditions
under which one can embed space-times with CTC into string/M-theory. In the
supergravity approximation of string theory/M-theory such embeddings were
considered [2-5]. Also various mechanisms which protect from CTC were
considered [2,6]. In the case of the G\"{o}del type space-times such rule
were proposed in [2] with the use of the holography.

The G\"{o}del space-time is an example of a homogenous solution with closed
time-like curves through every point. It is also non-singular with a
globally well-defined timelike Killing vector. In [3] the supersymmetric
extensions of the G\"{o}del space-time in five dimensions has been obtained
by using Killing spinors (in the supersymmetric case the Killing vectors are
obtained from covariantly constant Killing spinors). Another way of
obtaining\ G\"{o}del space-times in string theory/M-theory is the T or
S-dualizing a supersymmetric pp-wave solutions [2,4].

The G\"{o}del space-time has been obtained as an exact solution $M_{3}$ of
three-dimensional gravity coupled to a Maxwell-Chern-Simons (MCS) theory
[7]. It is also known that the four-dimensional G\"{o}del space-time has the
structure $M_{3}\times \mathbf{R}^{1}$. The gauge group of MCS is $U\left(
1\right) $. As turns out the 3-dimensional gauge theories with Chern-Simons
(CS) terms are realized in brane constructions in type IIB string theory
[8,9]. In [10] was shown that the 3-dimensional $\mathcal{N}=6$
supersymmetric $U\left( N\right) \times U\left( N\right) $ Chern-Simons
theory coupled to the matter is equivalent to the low-energetic theory on $N$
M2-branes at a $\mathbf{C}^{4}/\mathbf{Z}_{k}$ singularity and $k$ is a
level of the Chern-Simons theory. This observation follows from the fact
that the moduli space of the supersymmetric Chern-Simons is $\mathbf{C}^{4}/%
\mathbf{Z}_{k}$ which is the same as the moduli space for $N$ M2-branes
probing $\mathbf{C}^{4}/\mathbf{Z}_{k}$ singularity in M-theory. For $N>>1$
this low-energetic theory has a dual description in terms of M-theory on the
background $AdS_{4}\times S^{7}/\mathbf{Z}_{k}$. For $N/k$ fixed and $%
N\rightarrow \infty $ (the t'Hooft limit) the dual theory becomes type IIA
string theory on $AdS_{4}\times \mathbf{CP}^{3}$ background. From the other
side the 3-dimensional G\"{o}del space-time is obtained as the result of the
wrapping of M2-brans in the flux compactification in M-theory [11].

As turns out G\"{o}del and AdS metrics in (2+1) dimensions belong to the one
parameter family metrics [12]. In this family the Anti de Sitter (AdS) is
the boundary between space-times with CTC and without CTC. Thus it is
interesting to relate G\"{o}del-type solutions to AdS/CFT duality, in
particular, what closed timelike curves (CTC) mean in conformal field theory
(CFT). Some proposal relating causality region (without CTC) in the case of
the rotating black hole to unitarity of the dual CFT is given in [13]. The
AdS/CFT duality relate the string theory on 10-dimensional space-time $X$
(which is asymptotic to $AdS_{n}\times K_{10-n}$) to CFT defined on
conformal boundary of $AdS_{n}$. This duality is now well established and
was checked for different $n$. One can ask what will be changed in AdS/CFT
duality if AdS is replaced by one parameter family metrics of [12]. The
heterotic string propagating on deformed 3-dimensional AdS spaces were
considered in [14]. These backgrounds correspond to exactly marginal
deformations of the worldsheet conformal field theory. The string theory
models in Taub-NUT geometry with CTC's were considered in [15].

From the above one can see that the problem on causality is translated on
language of conformal field theory and/or string theory.

In this paper we consider G\"{o}del-type metrics in three dimensions from
the point of views of gauge theory (in section 2) and sigma model (in
section 3). We will be interested in conditions of appearing CTC's and their
behavior in the sigma models. Because gravity in 2+1 dimensions is expressed
in terms of gauge theory [16] so the gravity and MCS become the coupled
system of the two gauge theories: the first one is the non-abelian
corresponding to gravity and the second one is abelian corresponding to
Maxwell field. Thus CTC should have counterparts in such coupled gauge
theory. In section 2 we recall properties of gravity in (2+1)-dimensions as
the Chern-Simons theory and properties of the G\"{o}del-type metrics. In
this section we also give relation between coupling constants of the these
gauge theories in the case of the CTC. In section 3 we consider sigma-model
with target given by G\"{o}del-type space-time and show that exist
transformations which transform space-time with CTC on space-time without
CTC. Section 4 is devoted to conclusions. In Appendix we recall the symmetry
of G\"{o}del-type metrics and relation with AdS metric.

\section{G\"{o}del metrices and gauge theories in 3-dimensions}

The Hilbert-Eistein action with a cosmological constant $\Lambda $ in
(2+1)-dimensional space-time $M$ has the form:%
\begin{equation}
S[g_{\alpha \beta }]=\frac{1}{16\pi G}\int_{M}d^{3}x\sqrt{-g}\left(
R-2\Lambda \right) ,  \tag{2.1}
\end{equation}%
where $g_{\alpha \beta }$ is a metric on $M$. As is well-known one can go
from the metric $g_{\alpha \beta }$ to the first order forms $%
e^{a}=e_{\alpha }^{a}dx^{\alpha }$ and spin connections $\omega ^{a}=\omega
_{\alpha }^{a}dx^{\alpha }$ and as the result obtains the action $S$ in
terms of $e$ and $\omega $:%
\begin{gather}
S[e,\omega ]=\frac{1}{16\pi G}\int_{M}\left[ e^{a}\wedge \left( d\omega ^{a}+%
\frac{1}{2}\varepsilon _{abc}\omega ^{b}\wedge \omega ^{c}\right) \right. 
\notag \\
\left. +\frac{\Lambda }{6}\varepsilon _{abc}e^{a}\wedge e^{b}\wedge e^{c}%
\right] .  \tag{2.2}
\end{gather}%
The above action can be written as a sum of the Chern-Simons (CS) terms with 
$SL\left( 2,\mathbf{R}\right) $ gauge groups only in (2+1)-dimensions
[16,17]. For negative cosmological constant $\Lambda =-1/l^{2}$ the action
takes the form:%
\begin{equation}
S[A_{L},A_{R}]=k_{L}CS[A_{L}]-k_{R}CS[A_{R}],  \tag{2.3}
\end{equation}%
where $CS[A]$ is the Chern-Simons term:%
\begin{equation}
CS[A]=\frac{1}{4\pi }\int_{M}Tr\left( A\wedge dA+\frac{2}{3}A\wedge A\wedge
A\right)  \tag{2.4}
\end{equation}%
and $k_{L}+k_{R}=l/\left( 8G\right) $. The one-forms $A_{L,R}^{a}=\omega
^{a}\pm le^{a}$ are independent with values in $sl\left( 2,\mathbf{R}\right) 
$. In this way on the classical level Einstein gravity in (2+1) space-time
is the same as the CS theory with the gauge group $SL\left( 2,\mathbf{R}%
\right) \times SL\left( 2,\mathbf{R}\right) \approx SO\left( 2,2\right) $.
This gauge group is the symmetry of the 3-dimensional AdS space-time. For
this gauge theory one can add abelian gauge field $b_{\alpha }$ with Maxwell
action and abelian Chern-Simons term. The action for this system is:%
\begin{equation}
S_{MCS}[a]=\frac{1}{16\pi G}\int_{M}d^{3}x\left( \frac{\sqrt{-g}}{4}%
F^{\alpha \beta }F_{\beta \alpha }-\frac{k_{A}}{4\pi }\varepsilon ^{\alpha
\beta \gamma }b_{\alpha }F_{\beta \gamma }\right) ,  \tag{2.5}
\end{equation}%
where $F_{\alpha \beta }=\partial _{\alpha }b_{\beta }-\partial _{\beta
}b_{\alpha }$. Thus the gravity and MCS theory is the coupled gauge theory
with the gauge group $SO\left( 2,2\right) \times SO\left( 2\right) $ and the
action:%
\begin{gather}
S[A_{L},A_{R},b]=k_{L}CS[A_{L}]-k_{R}CS[A_{R}]  \notag \\
-k_{A}CS[b]+S[A_{L,R},db],  \tag{2.6}
\end{gather}%
where the last term is interaction part of the theory and is exactly the
Maxwell action for $b$:%
\begin{equation}
S[A_{L,R},db]=\frac{1}{4}\int db\wedge \ast db.  \tag{2.7}
\end{equation}%
The Hodge star $\ast $ is expressed in terms of connections $A_{L}$ and $%
A_{R}$. As was shown in [7] such the system has as a solution in the form of
the G\"{o}del-type metric and the Maxwell field $F=db$. This metric and the
gauge abelian field $b$ can be written as follows [18]:

\begin{gather}
ds^{2}=-\left( dt+\frac{4\Omega }{m^{2}}\sinh ^{2}\left( \frac{m\rho }{2}%
\right) d\phi \right) ^{2}  \notag \\
+d\rho ^{2}+\frac{1}{m^{2}}\sinh ^{2}\left( m\rho \right) d\phi ^{2}, 
\tag{2.8}
\end{gather}%
\begin{equation}
b\left( \rho \right) =\frac{4}{m^{2}}\left( \Omega ^{2}-1/l^{2}\right)
^{1/2}\sinh ^{2}\left( \frac{m\rho }{2}\right) d\phi ,  \tag{2.9}
\end{equation}%
where the cosmological constant $1/l$ is related to $\Omega $ and $m$ as
follows: $m^{2}-2\Omega ^{2}=2/l^{2}$. The vorticity $\Omega $ is related
with the Chern-Simons coupling $k_{A}$ : $\Omega =k_{A}/\left( 2\pi \right) $%
. Then we obtain relation between coupling constants and the parameter $m$:%
\begin{equation}
m^{2}=\frac{k_{A}^{2}}{2\pi ^{2}}+\frac{2}{\left( 8G\right) ^{2}\left(
k_{L}+k_{R}\right) ^{2}}.  \tag{2.10}
\end{equation}%
From the topological reasons the coupling constants $(k_{R}-k_{L})$ and $%
k_{A}$ are integers [17]. The metric (2.8) and the field (2.9) have the
following limits when $m\rightarrow 0$:%
\begin{equation}
ds^{2}=-\left( dt+\Omega \rho ^{2}d\phi \right) ^{2}+d\rho ^{2}+\rho
^{2}d\phi ^{2},  \tag{2.11}
\end{equation}%
\begin{equation}
b\left( \rho \right) =\left( \Omega ^{2}-1/l^{2}\right) ^{1/2}\rho ^{2}d\phi
.  \tag{2.12}
\end{equation}%
In this limit CTC's appear if $\rho >1/\Omega $.

The metric (2.8) can also be expressed as the one parameter family of the
3-dimensional metrics interpolating between the G\"{o}del metric and the AdS
metric [12]:%
\begin{gather}
ds_{\mu }^{2}=4a^{2}\left\{ -d\tau ^{2}+dr^{2}+\sinh ^{2}\left( r\right) 
\left[ 1+\left( 1-\mu ^{2}\right) \sinh ^{2}\left( r\right) \right] d\phi
^{2}\right.  \notag \\
\left. -2\mu \sinh ^{2}\left( r\right) d\tau d\phi \right\} ,  \tag{2.13}
\end{gather}%
where the parameter $\mu $ and the radius $a$ are related to vorticity $%
\Omega $ and $m$ in the following way: $\mu ^{2}=4\Omega ^{2}/m^{2}$ and $%
a^{2}=1/m^{2}$. The coordinates $\tau $ and $r$ are related to $t$ and $\rho 
$ by transformation: $t=2\tau /m$ and $\rho =2r/m$. The CTC appears if $%
g_{\phi \phi }\left( r\right) $ becomes the time-like: $g_{\phi \phi }\left(
r\right) <0$ which occurs for $r>r_{c}=\frac{1}{2}\ln \left( \frac{\mu +1}{%
\mu -1}\right) $ and $\mu >1$. As it is well-known the AdS metric and the G%
\"{o}del metric are obtained for $\mu ^{2}=1$ and for $\mu ^{2}=2$
respectively. Using (2.10) the parameter $\mu ^{2}$ is equal to:%
\begin{equation}
\mu ^{2}=\frac{2k_{A}^{2}}{k_{A}^{2}+\left( \pi /4G\right) ^{2}\left(
k_{L}+k_{R}\right) ^{-2}}.  \tag{2.14}
\end{equation}%
Thus the CTC appears if:%
\begin{equation}
k_{A}^{2}\left( k_{L}+k_{R}\right) ^{2}>\left( \pi /4G\right) ^{2}. 
\tag{2.15}
\end{equation}%
As one can see under the change of $k_{A}$ on $k_{L}+k_{R}$ the eq. (2.14)
and the condition (2.15) are invariant.

The AdS metric is obtained if: $k_{A}^{2}\left( k_{L}+k_{R}\right)
^{2}=\left( \pi /4G\right) ^{2}$ . In the case of the G\"{o}del metric ($\mu
^{2}=2$) one gets that $\left( k_{L}+k_{R}\right) ^{-2}=0$. It means that $%
\Lambda =0$ and $\Lambda $ is replaced by $\Omega $ given by $k_{A}$. In
this way we obtained condition on CTC expressed by the coupling constants of
the gauge theory. It is similiar to the condition on BPS states. The state
is BPS if a mass and a charge of the state are equal. In our case the CTC
disappears if (2.15) becomes equality what corresponds to the AdS metric.

\section{Sigma-model in the G\"{o}del type backgrounds}

In the conformal gauge on the worldsheet $\Sigma _{2}$ the bosonic part of
the action for sigma-model has the form:%
\begin{gather}
S=\frac{1}{2\pi }\int_{\Sigma _{2}}d^{2}z\left( G_{MN}+B_{MN}\right)
\partial X^{M}\overline{\partial }X^{N}  \notag \\
+\frac{1}{4\pi }\int_{\Sigma _{2}}\Phi \left( X\right) R^{\left( 2\right)
}d^{2}z,  \tag{3.1}
\end{gather}%
where $X$ is the mapping from $\Sigma _{2}$ to the manifold $M_{n}$ with the
background metric $G_{MN}$, antisymmetric form $B_{MN}$ and dilaton field $%
\Phi $. The scalar curvature of $\Sigma _{2}$ is denoted by $R^{\left(
2\right) }$. In the case when the background is given by the metric (2.8)
and the field $X$ is parametrized as follows: $X=\left( \tau ,r,\phi \right) 
$, then the action takes the form: 
\begin{gather}
S[\tau ,r,\phi ]=\frac{4a^{2}}{2\pi }\int_{\Sigma _{2}}d^{2}z\left[
-\partial \tau \overline{\partial }\tau +\partial r\overline{\partial }%
r\right. +\left( \sinh ^{2}\left( r\right) +\left( 1-\mu ^{2}\right) \sinh
^{4}\left( r\right) \right) \partial \phi \overline{\partial }\phi   \notag
\\
\left. -\mu \sinh ^{2}\left( r\right) \left( \partial \phi \overline{%
\partial }\tau +\overline{\partial }\phi \partial \tau \right) \right] +%
\frac{1}{4\pi }\int_{\Sigma _{2}}\Phi R^{\left( 2\right) }d^{2}z,  \tag{3.2}
\end{gather}%
with the background symmetry generated by the algebra $so\left( 2\right)
\times sl\left( 2,\mathbf{R}\right) $. If one considers the above background
in the sigma model, then the consistency conditions need a non-zero B-field
in order to solve the sigma-model beta function equations. However let us
assume here that the field B is equal to zero. As one can notice this action
is invariant under shift symmetry in the variables $\tau $ and $\phi $.
Since this symmetry is abelian we can rewrite the action as follows [19]: 
\begin{gather}
S[\tau ,r,\phi ]=  \notag \\
\frac{1}{2\pi }\int_{\Sigma _{2}}d^{2}z\left( 
\begin{array}{ccc}
\partial \tau  & \partial \phi  & \partial r%
\end{array}%
\right) \left( 
\begin{array}{cc}
E & 0 \\ 
0 & 4a^{2}%
\end{array}%
\right) \left( 
\begin{array}{c}
\overline{\partial }\tau  \\ 
\overline{\partial }\phi  \\ 
\overline{\partial }r%
\end{array}%
\right)   \notag \\
+\frac{1}{4\pi }\int_{\Sigma _{2}}\Phi R^{\left( 2\right) }d^{2}z,  \tag{3.3}
\end{gather}%
where the matrix $E$ is equal to:%
\begin{equation}
E=4a^{2}\left( 
\begin{array}{cc}
-1 & -\mu \sinh ^{2}r \\ 
-\mu \sinh ^{2}r & u\left( r\right) \sinh ^{2}r%
\end{array}%
\right)   \tag{3.4}
\end{equation}%
and $u\left( r\right) =1+\left( 1-\mu ^{2}\right) \sinh ^{2}r$. We can
deform this conformal field theory by exactly marginal operators. In the
case of the toroidal background these operators correspond to a one
parameter families of $O\left( d,d,\mathbf{R}\right) $ rotations, where $d$
is a number of the abelian isometries. In the considered model $d=2$ thus on
the target space side the exactly marginal operators correspond to a one
parameter families of $O\left( 2,2,\mathbf{R}\right) $ rotations. The group $%
O\left( 2,2,\mathbf{R}\right) $ acts on the background represented by a
matrix $E=G+B$ in the following way:%
\begin{equation}
g\left( E\right) =\left( aE+b\right) \left( cE+d\right) ^{-1},  \tag{3.5}
\end{equation}%
where $g\in O\left( 2,2,\mathbf{R}\right) $ has the matrix representation:%
\begin{equation}
g=\left( 
\begin{array}{cc}
a & b \\ 
c & d%
\end{array}%
\right)   \tag{3.6}
\end{equation}%
and the matrices $a,b,c$ and $d$ fulfill relations: $%
a^{T}d+c^{T}a=b^{T}d+d^{T}b=0$, $a^{T}d+c^{T}b=I_{2}$. The maximal compact
subgroup of $O\left( 2,2,\mathbf{R}\right) $ is $O\left( 2\right) \times
O\left( 2\right) $. The embedding $e:O\left( 2\right) \times O\left(
2\right) \rightarrow O\left( 2,2,\mathbf{R}\right) $ has the form:%
\begin{equation}
e\left( r_{1},r_{2}\right) =\frac{1}{2}\left( 
\begin{array}{cc}
r_{1}+r_{2} & r_{1}-r_{2} \\ 
r_{1}-r_{2} & r_{1}+r_{2}%
\end{array}%
\right) ,  \tag{3.7}
\end{equation}%
where $\left( r_{1},r_{2}\right) \in O\left( 2\right) \times O\left(
2\right) $. This subgroup depends on two angles $\alpha _{1}$ and $\alpha
_{2}$ which parametrize the element $\left( r_{1}\left( \alpha _{1}\right)
,r_{2}\left( \alpha _{2}\right) \right) $ in the standard way:%
\begin{equation}
r_{i}\left( \alpha _{i}\right) =\left( 
\begin{array}{cc}
\cos \alpha _{i} & \sin \alpha _{i} \\ 
-\sin \alpha _{i} & \cos \alpha _{i}%
\end{array}%
\right)   \tag{3.8}
\end{equation}%
and $i=1,2$. Thus the embedding $e$ has the form:%
\begin{equation}
e\left( r_{1},r_{2}\right) =\left( 
\begin{array}{cc}
r\left( \alpha \right) \cos \beta  & r\left( \alpha \right) \varepsilon \sin
\beta  \\ 
r\left( \alpha \right) \varepsilon \sin \beta  & r\left( \alpha \right) \cos
\beta 
\end{array}%
\right) \equiv e_{\left( \alpha ,\beta \right) },  \tag{3.9}
\end{equation}%
where the angles $\alpha $ and $\beta $ are related to $\alpha _{1}$ and $%
\alpha _{2}$ as follows: $\alpha =\left( \alpha _{1}+\alpha _{2}\right) /2$, 
$\beta =\left( \alpha _{1}-\alpha _{2}\right) /2$ the matrices $r\left(
\alpha \right) $ and $\varepsilon $ are:%
\begin{equation}
r\left( \alpha \right) =\left( 
\begin{array}{cc}
\cos \alpha  & \sin \alpha  \\ 
-\sin \alpha  & \cos \alpha 
\end{array}%
\right) \text{ and }\varepsilon =\left( 
\begin{array}{cc}
0 & 1 \\ 
-1 & 0%
\end{array}%
\right) .  \tag{3.10}
\end{equation}%
The action of this maximal compact group on $E$ depends on two parameters $%
\alpha $ and $\beta $ and gives new background matrix $\widetilde{E}$:%
\begin{equation}
e_{\left( \alpha ,\beta \right) }\left( E\right) \equiv \widetilde{E}%
=r\left( \alpha \right) \left[ E+\varepsilon \tan \beta \right] \left[
\varepsilon E\tan \beta +I\right] ^{-1}r^{T}\left( \alpha \right) . 
\tag{3.11}
\end{equation}%
In the considered case the background field $B$ is vanishing thus the matrix 
$E=G$ and the metric $G$ is given by (2.13). Under the deformations related
to the maximal compact subgroup $O\left( 2\right) \times O\left( 2\right) $
the matrix $\widetilde{E}$ has the entries:%
\begin{gather}
\widetilde{E}_{11}=W_{\beta }\left( r\right) \left[ \left( 1-\mu ^{2}\right)
\sin ^{2}\alpha \sinh ^{4}r\right.   \notag \\
\left. +\left( \sin ^{2}\alpha -\mu \sin 2\alpha \right) \sinh ^{2}r-\cos
^{2}\alpha \right] ,  \tag{3.12}
\end{gather}%
\begin{gather}
\widetilde{E}_{12}=\frac{W_{\beta }\left( r\right) }{2}\left[ \left( 1-\mu
^{2}\right) \sin 2\alpha \sinh ^{4}r\right.   \notag \\
\left. +\left( \sin 2\alpha -2\mu \cos 2\alpha \right) \sinh ^{2}r+\sin
2\alpha +2\widetilde{b}\right] ,  \tag{3.13}
\end{gather}%
\begin{gather}
\widetilde{E}_{21}=\frac{W_{\beta }\left( r\right) }{2}\left[ \left( 1-\mu
^{2}\right) \sin 2\alpha \sinh ^{4}r\right.   \notag \\
\left. +\left( \sin 2\alpha -2\mu \cos 2\alpha \right) \sinh ^{2}r+\sin
2\alpha -2\widetilde{b}\right] ,  \tag{3.14}
\end{gather}%
\begin{gather}
\widetilde{E}_{22}=W_{\beta }\left( r\right) \left[ \left( 1-\mu ^{2}\right)
\cos ^{2}\alpha \sinh ^{4}r\right.   \notag \\
\left. +\left( \cos ^{2}\alpha -\mu \sin 2\alpha \right) \sinh ^{2}r-\sin
^{2}\alpha \right] ,  \tag{3.15}
\end{gather}%
where the functions $W_{\beta }\left( r\right) $, $\widetilde{b}\left(
r\right) $ are equal to: 
\begin{gather}
W_{\beta }\left( r\right) =\frac{4a^{2}\left( 1+\tan ^{2}\beta \right) }{%
1-4a^{4}\tan ^{2}\beta \sinh ^{2}\left( 2r\right) },\text{ }  \notag \\
\widetilde{b}=\frac{1}{8}\left( 1+4a^{2}\sinh ^{2}(2r)\right) \sin \left(
2\beta \right) .  \tag{3.16}
\end{gather}%
One can read off from the above formulas that the new background has the
metric $\widetilde{G}$ given by :%
\begin{equation}
\widetilde{G}_{11}=\widetilde{E}_{11},\text{ \ \ }\widetilde{G}_{22}=%
\widetilde{E}_{22}  \tag{3.17}
\end{equation}%
and 
\begin{gather}
\widetilde{G}_{12}\left( \alpha ,\beta \right) =\frac{W_{\beta }\left(
r\right) }{2}\left[ \left( 1-\mu ^{2}\right) \sin 2\alpha \sinh ^{4}r\right. 
\notag \\
\left. +\left( \sin 2\alpha -2\mu \cos 2\alpha \right) \sinh ^{2}r+\sin
2\alpha \right] ,  \tag{3.18}
\end{gather}%
with the antisymmetric two form $\widetilde{B}$ :%
\begin{equation}
\widetilde{B}=\frac{W_{\beta }\left( r\right) }{8}\left( 1+4a^{2}\sinh
^{2}(2r)\right) \sin \left( 2\beta \right) d\tau \wedge d\phi \text{.} 
\tag{3.19}
\end{equation}%
For $\alpha =\beta =0$ we obtain the initial matrix $E$. The function $%
W_{\beta }\left( r\right) $ is finite for $\beta =\pi /2$ and has the value $%
W_{\pi /2}\left( r\right) =-a^{-2}\sinh ^{-2}\left( 2r\right) $. In the case
when $\mu ^{2}>1$ the forms of the $\widetilde{G}_{11}$, $\widetilde{G}_{12}$
and $\widetilde{G}_{22}$ are:%
\begin{equation}
\widetilde{G}_{11}\left( \alpha ,\beta \right) =\left( \mu ^{2}-1\right)
W_{\beta }\left( r\right) h\left( R\right) \sin ^{2}\alpha ,  \tag{3.20}
\end{equation}%
\begin{equation}
\widetilde{G}_{12}\left( \alpha ,\beta \right) =\frac{1}{2}\left( \mu
^{2}-1\right) W_{\beta }\left( r\right) g\left( R\right) \sin 2\alpha , 
\tag{3.21}
\end{equation}%
\begin{equation}
\widetilde{G}_{22}\left( \alpha ,\beta \right) =\left( \mu ^{2}-1\right)
W_{\beta }\left( r\right) f\left( R\right) \cos ^{2}\alpha ,  \tag{3.22}
\end{equation}%
where:%
\begin{gather}
h\left( R\right) =\left( \frac{1-2\mu \cot \alpha }{2\left( \mu
^{2}-1\right) }\right) ^{2}-\frac{\cot ^{2}\alpha }{\mu ^{2}-1}  \notag \\
-\left( R-\frac{1-2\mu \cot \alpha }{2\left( \mu ^{2}-1\right) }\right) ^{2},
\tag{3.23}
\end{gather}%
\begin{gather}
g\left( R\right) =\left( \frac{1-2\mu \cot 2\alpha }{2\left( \mu
^{2}-1\right) }\right) ^{2}-1  \notag \\
-\left( R-\frac{1-2\mu \cot 2\alpha }{2\left( \mu ^{2}-1\right) }\right)
^{2},  \tag{3.24}
\end{gather}%
\begin{gather}
f\left( R\right) =\left( \frac{1-2\mu \tan \alpha }{2\left( \mu
^{2}-1\right) }\right) ^{2}-\frac{\tan ^{2}\alpha }{\mu ^{2}-1}  \notag \\
-\left( R-\frac{1-2\mu \tan \alpha }{2\left( \mu ^{2}-1\right) }\right) ^{2}
\tag{3.25}
\end{gather}%
and $R=\sinh ^{2}r$. One can notice that for $\alpha =\pi /2$ the metric
components are following:%
\begin{equation}
\widetilde{G}_{11}\left( \pi /2,\beta \right) =\widetilde{G}_{22}\left(
0,\beta \right) =\frac{1}{4a^{2}}W_{\beta }\left( r\right) G_{22}, 
\tag{3.26}
\end{equation}%
\begin{equation}
\widetilde{G}_{22}\left( \pi /2,\beta \right) =\widetilde{G}_{11}\left(
0,\beta \right) =\frac{1}{4a^{2}}W_{\beta }\left( r\right) G_{11}, 
\tag{3.27}
\end{equation}%
\begin{equation}
\widetilde{G}_{12}\left( \pi /2,\beta \right) =-\frac{1}{4a^{2}}W_{\beta
}\left( r\right) G_{12}.  \tag{3.28}
\end{equation}%
It means that under $\alpha $-rotation (for $\alpha =\pi /2$) the time and
space coordinates are changing their positions. In such rotated background
there is a nonvanishing two-form $\widetilde{B}$ which depends on the second
rotation angle $\beta $. Moreover for $\alpha =\pi /2$ and $\beta =\pi /2$
the deformed background is conformal equivalent to the initial with the
changing time and space coordinates with the vanishing two-form and the
conformal factor is equal to $W_{\pi /2}\left( r\right) $. One can also
notice the following relation:%
\begin{equation}
\widetilde{G}_{11}\left( \alpha ,\beta \right) +\widetilde{G}_{22}\left(
\alpha ,\beta \right) =\left( \mu ^{2}-1\right) W_{\beta }\left( r\right)
p\left( R\right) ,  \tag{3.29}
\end{equation}%
where%
\begin{equation}
p\left( R\right) =-R^{2}-\frac{2\mu \sin 2\alpha -1}{\mu ^{2}-1}R+\frac{1}{%
\mu ^{2}-1}.  \tag{3.30}
\end{equation}%
Since the group $O(2)\times O(2)$ has two generators thus the exactly
marginal operators correspond to two independent one-parametric families
related to the rotation in $\U{3b1} $ angle and the rotation in $\U{3b2} $
angle, respectively. This independence of the two families exactly marginal
operators is reflected in the factorized form of the transformed background
(3.19-3.22) where the factors depend on $\U{3b1} $ and $\U{3b2} .$\emph{\  \
\  }

In the background given by (3.19)-(3.22) CTC appears if $\widetilde{G}_{22}$
becomes negative whereas $\widetilde{G}_{11}$ remains also negative. Thus
CTC appears if: $f\left( R\right) <0$ and $W_{\beta }\left( r\right) >0$ or $%
f\left( R\right) >0$ and $W_{\beta }\left( r\right) <0$ under condition that 
$\widetilde{G}_{11}<0$. This last condition implicated that: $W_{\beta
}\left( r\right) <0$ and $h\left( R\right) >0$ or $W_{\beta }\left( r\right)
>0$ and $h\left( R\right) <0$. The function $W_{\beta }\left( r\right) >0$
for $\sinh ^{2}\left( 2r\right) <\left( \cot ^{2}\beta \right) /\left(
4a^{2}\right) $. It is easy to see that $f\left( R\right) >0$ if $R\in
\left( R_{-}^{f},R_{+}^{f}\right) \equiv I^{f}$ where:%
\begin{equation}
R_{\pm }^{f}=\frac{1}{2\left( \mu ^{2}-1\right) }\left( 1-2\mu \tan \alpha
\pm 2\sqrt{\left( \tan \alpha -\mu /2\right) ^{2}-\left( \mu ^{2}-1\right) /4%
}\right)   \tag{3.31}
\end{equation}%
and $f\left( R_{\pm }^{f}\right) =0$. One can see that $R_{\pm }^{f}$ are
real if $\tan \alpha \in (0,\nu _{-}/2]\cup \lbrack \nu _{+}/2,+\infty
)\equiv D_{f}$ where $\nu _{\pm }=\mu \pm \sqrt{\mu ^{2}-1}$. In the case
when $\tan \alpha \in \left( \nu _{-}/2,\nu _{+}/2\right) $ the function $%
f\left( R\right) $ is always negative. Let us consider the case with $\U{3b2}
=\U{3c0} /2$. This choice of $\U{3b2} $ is a matter of convenience and
simplifies computations. For other values of $\U{3b2} $ the qualitative
picture of our considerations does not change. Thus in our case $W_{\pi
/2}\left( r\right) <0$ and CTC exists only when $f\left( R\right) >0$ and $%
h\left( R\right) >0$. The last inequality holds for $R\in \left(
R_{-}^{h},R_{+}^{h}\right) \equiv I^{h}$ where: 
\begin{equation}
R_{\pm }^{h}=\frac{1}{2\left( \mu ^{2}-1\right) }\left( 1-2\mu \cot \alpha
\pm 2\sqrt{\left( \cot \alpha -\mu /2\right) ^{2}-\left( \mu ^{2}-1\right) /4%
}\right)   \tag{3.32}
\end{equation}%
and $h\left( R_{\pm }^{h}\right) =0$. These real roots exist if $\tan \alpha
\in (0,2\nu _{-}]\cup \lbrack 2\nu _{+},+\infty )\equiv D_{h}$. For $\tan
\alpha \in \left( 2\nu _{-},2\nu _{+}\right) $ the function $h\left(
R\right) $ is always negative. Thus the real roots $R_{\pm }^{f}$ and $%
R_{\pm }^{h}$ exist if $\tan \alpha \in D_{f}\cap D_{h}=(0,\nu _{-}/2]\cup %
\left[ \nu _{+}/2,2\nu _{-}\right] \cup \lbrack 2\nu _{+},+\infty )$.

It follows that the CTC's appear iff $I^{f}\cap I^{h}\neq \varnothing $.
This relation is valid if exists such $R_{0}>0$ that $f\left( R_{0}\right)
=h\left( R_{0}\right) >0$. As one can easily find that $R_{0}=-\left( \mu
\sin 2\alpha \right) ^{-1}$ and this radius has to be greater than zero so
we get that $\alpha \in \left( \pi /2,\pi \right) $. The value of the
function $f$ in the point $R_{0}$ is:%
\begin{equation}
f\left( R_{0}\right) =\frac{1}{\left( \mu ^{2}-1\right) \sin ^{2}\left(
2\alpha \right) }Z\left( \alpha \right) ,  \tag{3.33}
\end{equation}%
where:%
\begin{gather}
Z\left( \alpha \right) =4\left( \mu ^{3}-1\right) \cos ^{4}\alpha -\left(
\mu ^{2}-1\right) \sin ^{4}\alpha  \notag \\
-\left( 4\mu \sin ^{2}\alpha -\cos ^{2}\alpha \right) \sin 2\alpha +\mu \sin
^{2}\left( 2\alpha \right) -\cos ^{2}\alpha  \tag{3.34}
\end{gather}%
and $\alpha \in \left( \pi /2,\pi \right) $. The sign of the function $Z$
depends on the value of $\mu $. As the example we consider the G\"{o}del
metric ($\mu ^{2}=2$). The function $Z\left( \alpha \right) $ is negative
for $\alpha \in \lbrack 1.57,$ $1.8]$ (in radian) and becomes positive for $%
\alpha >1.801$. Another considered example is for $\mu ^{2}=\left(
1.5\right) ^{2}$ in this case we get that $Z\left( \alpha \right) $ is
negative for $\alpha \in \lbrack 1.57,$ $1.67]$. Thus the transformation\
(3.11) with $\alpha $ from these intervals (corresponding to $\mu ^{2}=2$
and $\mu ^{2}=\left( 1.5\right) ^{2}$) takes the targets with CTC to the
targets where the CTC is vanishing.

It means that in the considered case one can make starting with the G\"{o}%
del-type metrics the transformation belonging to $O(2)\times O(2)$ which
changes the causal structure. In our case other transformations from $O(2,2,%
\mathbf{R})/(O(2)\times O(2))$ exist which form the moduli space for the
CFT. The group $O(2,2,\mathbf{R})$ (in general case $O(d,d,\mathbf{R})$
group) has interpretation as the duality group, where the T-duality is
realized by $O(2,2,\mathbf{Z})$ subgroup. In general case where the number
of isometries is $d$ the group $O(2,2,\mathbf{R})$ is replaced by $O(d,d,%
\mathbf{R})$. This group is the structure group of the generalized tangent
bundle, which combines the tangent and cotangent bundles of a $d$%
-dimensional manifold in generalized geometry [20, 21]. This suggests the
use of generalized geometry to describe causal structure.

\section{Conclusions}

We considered space-times with the closed time-like curves. These
space-times were represented by the G\"{o}del-type metrics in three
dimensions. We interested in the conditions of appearing CTC's and their
behavior from the point of views gauge theory and conformal theory. In the
section 2 we obtained condition on CTC's expressed by the coupling constants
of the gauge theory. In the section 3 we started with the sigma-model with
the target space given by G\"{o}del-type spacetimes. Assuming that this
model realize the conformal field theory we used exactly marginal operators
represented by $O\left( 2,2\right) $ on the target space side in order to
get new targets which are not connected by target space diffeomorphismes.
This assumption is justified by the results that G\"{o}del-type spacetimes
can be considered as the backgrounds for string propagation [4, 11, 14]. We
obtained that in some of these new targets CTC's are vanishing whereas in
other they still exist. Thus the exactly marginal operators act on the
causal structure of the target. The transformed backgrounds have been here
shown in the explicit forms (3.17 -3.19). One can then see the analogy of
our result with the relation between the mirror symmetry and the $N=2$
superconformal field theory realized by the sigma-model with the Calabi-Yau
three-fold as a target. Note that the exactly marginal operators in the
superconformal field theory correspond to the harmonic (1,1)-forms and
(2,1)-forms. These forms are responsible for the deformation of the "size"
and the "shape" of this three-fold, respectively. In our case the exactly
marginal operators act on causal structure. We proved that some causal
spacetimes have counterparts in which causal protection is violated by
CTC's. From the other side the causal structure is related to time but on
the quantum mechanical side the operator corresponding to time does not
exist [22]. Also in approach to quantization of gravity based on Wheeler-De
Witt equation the time is vanishing as a consequence of the Hamiltonian
constraint. We used the transformations from the conformal field theory,
which is a starting point for quantization of the string theory, considered
in this case as a perturbative quantum theory of gravitation. Thus these
transformations represent the quantum effects related to the deformation of
the causal structure of the target.

Here we only considered the G\"{o}del-type metrics in (2+1) dimensions. We\
chose this example because it appears as the background for strings
propagation. Thus we applied the transformations used by the conformal field
theory.\emph{\ }The second reason for considering this background is that it
is (as the solution of equations of the general relativity) non-singular in
the curvature tensor. Although this solution does not describe our universe
it points out problems with the chronology in general relativity as it is
well-known from long time.

\section{Appendix}

Relations between one-forms $e$ and metric $g$ are:%
\begin{equation}
g_{\alpha \beta }=e_{\alpha }^{a}e_{\beta }^{b}\eta _{ab},  \tag{A.1}
\end{equation}%
where Minkowski metric $\eta _{ab}=diag\left( -1,1,1\right) $. The measure $%
\sqrt{-g}d^{3}x$ is expressed by $e$ as follows:%
\begin{equation}
\sqrt{-g}d^{3}x=\frac{1}{6}\varepsilon _{abc}e^{a}\wedge e^{b}\wedge e^{c}. 
\tag{A.2}
\end{equation}%
The vierbains $e^{a}$ are expressed by connections $A_{L,R}$ as follows: $%
e^{a}=\left( A_{L}^{a}-A_{R}^{a}\right) /\left( 2l\right) .$

The metric (2.13) has the four Killing vectors $\xi _{\left( 1\right)
},...,\xi _{\left( 4\right) }$ which span the $so\left( 2\right) \times
sl\left( 2,\mathbf{R}\right) $ algebra:%
\begin{equation*}
\left[ \xi _{\left( 1\right) },\xi _{\left( r\right) }\right] =0,
\end{equation*}%
\begin{eqnarray}
\left[ \xi _{\left( 2\right) },\xi _{\left( 3\right) }\right] &=&-\frac{1}{a}%
\xi _{\left( 2\right) },\text{ \ }\left[ \xi _{\left( 2\right) },\xi
_{\left( 4\right) }\right] =+\frac{1}{a}\xi _{\left( 3\right) },  \notag \\
\text{ \ }\left[ \xi _{\left( 3\right) },\xi _{\left( 4\right) }\right] &=&-%
\frac{1}{a}\xi _{\left( 4\right) }.  \TCItag{A.3}
\end{eqnarray}%
The concrete forms of those vectors depend on the choose coordinates in the
metric. In the above notation the abelian part $so\left( 2\right) $ is
spanned by $\xi _{\left( 1\right) }$. As one can shown these metrics are
solutions of the Einstein equations:%
\begin{equation}
R_{\alpha \beta }-\frac{1}{2}g_{\alpha \beta }R-\Omega ^{2}g_{\alpha \beta
}=T_{\alpha \beta },  \tag{A.4}
\end{equation}%
with the energy-momentum tensor $T_{\alpha \beta }$ is given by:%
\begin{equation}
T_{\alpha \beta }=\left( 4\Omega ^{2}-m^{2}\right) g_{\alpha 0}g_{\beta 0} 
\tag{A.5}
\end{equation}%
and $\alpha ,\beta =0,1,2$. In the case when $m^{2}=4\Omega ^{2}$ the
space-time becomes the AdS with the metric:%
\begin{gather}
ds_{\left( AdS\right) }^{2}=-\left( dt+\frac{1}{\Omega }\sinh ^{2}\left(
\Omega \rho \right) d\phi \right) ^{2}+d\rho ^{2}  \notag \\
+\frac{1}{4\Omega ^{2}}\sinh ^{2}\left( 2\Omega \rho \right) d\phi ^{2}. 
\tag{A.6}
\end{gather}%
In order to get standard form we have to make the following change of
coordinates $\sigma =\Omega t$, \ $\psi =\Omega t-\phi $ and $\widetilde{%
\rho }=\Omega \rho $. As the result one obtains:%
\begin{equation}
ds_{\left( AdS\right) }^{2}=\frac{1}{\Omega ^{2}}\left( -\cosh ^{2}\left( 
\widetilde{\rho }\right) d\sigma ^{2}+d\widetilde{\rho }^{2}+\sinh
^{2}\left( \widetilde{\rho }\right) d\psi ^{2}\right) .  \tag{A.7}
\end{equation}

\section{References}

[1] S.W. Hawking, \textit{The Chronology Protection Conjecture},\ Phys. Rev.
D46 (1992) 603.

[2] E. K. Boyda, S. Ganguli, P. Ho\v{r}ava, and U. Varadarajan, \textit{%
Holographic protection of chronology in universes of the G\"{o}del type},
Phys. Rev. D67 (2003) 106003, hep-th/0212087.

[3] C. A. R. Herdeiro, \textit{Spinning deformations of the D1-D5 system and
a geometric resolution of closed timelike curves},\ Nucl. Phys. B665 (2003)
189 [arXiv:hep-th/0212002].

[4] T. Harmark and T. Takayanagi, \textit{Supersymmetric Goedel universes in
string theory},\ Nucl. Phys. B662 (2003) 3 [arXiv: hep-th/0301206].

[5] J. P. Gauntlett, J. B. Gutowski, C. M. Hull, S. Pakis, and H. S. Reall, 
\textit{All supersymmetric solutions of minimal supergravity in five
dimensions}, Class.Quant.Grav. 20 (2003) 4587-4634 [arXiv: hep-th/0209114].

[6] N. Drukker, \textit{Supertube domain-walls and elimination of closed
time-like curves in string theory}, hep-th/0404239.

[7] M. Banados, G. Barnich, G. Compere, A. Gomberoff, \textit{Three
dimensional origin of G\"{o}del spacetimes and black holes,} Phys.Rev. D73
(2006) 044006, [hep-th/0512105].

[8] T. Kitao, K. Ohta and N. Ohta, \textit{Three-dimensional gauge dynamics
from brane configurations with (p,q)-fivebrane}, Nucl. Phys. B 539, 79
(1999) [arXiv:hepth/9808111].

[9] O. Bergman, A. Hanany, A. Karch and B. Kol, \textit{Branes and
supersymmetry breaking in 3D gauge theories}, JHEP 9910, 036 (1999)
[arXiv:hep-th/9908075].

[10] O. Aharony, O. Bergman, D. L. Jafferis, and J. Maldacena, \textit{N=6
superconformal Chern-Simons-matter theories, M2-branes and their gravity
duals, }JHEP 0810, 091 (2008) [arXiv:0806.1218].

[11] T. S. Levi, J. Raeymaekers, D. Van den Bleeken, W. Van Herck, B.
Vercnocke, \textit{Godel space from wrapped M2-branes,} [ arXiv:0909.4081].

[12] M. Rooman and P. Spindel, \textit{Goedel metric as a squashed anti-de
Sitter geometry}, Class. Quant. Grav. 15 (1998) 3241,
[arXiv:gr-qc/9804027].\ 

[13] L. Dyson, \textit{Chronology Protection in String Theory,}\ JHEP 0403,
024 (2004), [aXiv: hep-th/0302052].

[14] D. Israel, \textit{Quantization of heterotic strings in a G\"{o}%
del/Anti de Sitter spacetime and chronology protection,}\ JHEP 0401, 042
(2004), [arXiv: hep-th/0310158].

[15] C. V. Johnson and H. G. Svendsen,\textit{\ An exact string theory model
of closed time-like curves and cosmological singularities}, Phys. Rev. D70
(2004) 126011, [arXiv: hep-th/0405141].

[16] E. Witten, (2+1)-\textit{Dimensional Gravity as an Exactly Soluble
System}, Nucl. Phys. B311 (1988) 46.

[17] E. Witten, \textit{Three-dimensional Gravity Reconsidered},
[arXiv:0706.3359].

[18] M. J. Rebou\c{c}as and J. Tiomno, \textit{Homogeneity of Riemmannian
space-times of the G\"{o}del type, }Phys. Rev. \textbf{D28}, 1251 (1983).

[19] A. Giveon, M. Porrati, E. Rabinovici, \textit{Target Space Duality in
String Theory,}\ Phys.Rept. 244 (1994) 77-202, [arXiv: hep-th/9401139].

[20] N. Hitchin, \textit{Generalized Calabi-Yau manifolds}, Quart. J. Math.
Oxford Ser. 54 (2003) 281 [arXiv:math.dg/0209099].

[21] M. Gualtieri, \textit{Generalized Complex Geometry}, Oxford University
DPhil thesis (2004) [arXiv:math.DG/0401221].

[22] C. J. Isham, \textit{Canonical Quantum Gravity and the Problem of Time,}
[arXiv: gr-qc/9210011].

\end{document}